# How to benchmark: the Measure-Explain-Test-Improve loop

GABRIEL SCHERER, Picube, INRIA, France and IRIF, Université Paris Cité, France

I would like to share recommendations on how to do performance benchmarks for the purpose of computer science research evaluation. Research in my field (programming language research) often involves performance considerations, but it is typically not the main tool used to evaluate our research (typically we evaluate via formal statements and their proofs, experience writing large or interesting examples, or systematic comparison of expressivity, feature set, etc.). My impression is that, as a result, we tend to not do our performance evaluation very well.

In the present document I will try to explain a methodology to do benchmarking correctly (I hope!). People with no former benchmarking experience should be able to build solid performance evaluation as part of their research. I explain the justification for each aspect along the way.

## 1 Introduction

The main idea is that a performance evaluation is an iterative loop, not a one-step process. When we write software we typically run an iterative loop where we implement something, test it carefully, refine it, then move to the next step, test it carefully, refine it, and test and refine and test and refine until we eventually feel confident *enough* that the result is doing what it should. Benchmarks are similar, you refine them iteratively, fixing mistakes along the way, until you reach a state where you feel confident in your results. If you don't take the time to evaluate each step carefully and to iterate, the results will be terrible. In other words:

> There is a *testing* process for benchmarks; untested results can be assumed to be wrong.

Be prepared to spend a significant amount of time on your benchmarks. In my experience, doing a performance evaluation well typically requires *weeks* of work. In the specific case of a small but tricky data-structure implementation, I have observed that a careful performance evaluation took about the same amount of time and effort as proving the core of the data-structure correct in Rocq.

*Acknowledgments.* This is a topic that I have learned by doing, over the course of several research projects where I participated in performance evaluations, notably with my colleagues Guillaume Munch-Maccagnoni and Basile Clément; I learned much from their expertise and methodology.

Thanks are also due to Bob Atkey and Artem Pelenitsyn for providing advice and references.

Finally I must thank the OOPSLA'26 authors whose benchmarks motivated me to write this.

## 2 Related works

When I started doing benchmarks for real I looked for guides on how to do benchmarking. The document I found would tell *horror stories* about fields where reasoning about performance is *really very difficult* and it is much easier than you might think to shoot you in the foot. They also tend to give very detailed advice on how to setup a benchmarking machine: it has to be air-gapped; it must not have opened a web browser in the last forty days; you need to reboot the machine and wait a random number of hours between each iteration of the program under measurement; and you must use their scary scripts that applies 12 different mitigations against sources of measurement noise coming from your operating system or hardware (... but it only works on PowerPC machines).

I am sure that doing all of this is very important in problem domains where the performance differences being discussed are very small are very sensitive. On the other hand, I run my own benchmarks on my personal machine, and sometimes I even *check my email* while they are running!

Author's Contact Information: Gabriel Scherer, Picube, INRIA, Paris, France and IRIF, Université Paris Cité, France, gabriel.scherer@inria.fr.

In my experience running benchmarks like normal people run programs works well enough for the most part, and the recommendations in the present document is not about eliminating measurement noise.

Your first enemy is the human in the loop, that is *yourself*. You are going to make a ton of mistakes in the benchmark setup and make up a bunch of completely wrong explanations for the results observed. The recommendations here protect against this. Once you end up in a situation where the main source of benchmarking error are cosmic rays, congratulations, you evolved beyond this guide.

With that said, I would definitely recommend the following previous work on how to benchmark:

- "Producing wrong data without doing anything obviously wrong!", Mytkowicz, Diwan, Hauswirth, and Sweeney [2009]: A scary story about how different Unix usernames can result in a 10% difference in performance results. A section at the end, 7.2 on "using causal analysis", is similar in spirit to the main recommendations of the present document, the Measure-Explain-Test subsequence. We agree on how to produce good performance explanations, but it does not emphasize the iterative process of fixing benchmarking errors.
- "STABILIZER: statistically sound performance evaluation", Curtsinger and Berger [2013]: A complex system to eliminate many sources of environmental noises in your benchmarked programs.
- "Virtual Machine Warmup Blows Hot and Cold", Barrett, Bolz-Tereick, Killick, Mount, and Tratt [2017]: Scary battle stories of trying to come to valid conclusion about the performance of Just-In-Time language implementations, which are systems that are precisely designed to change their performance profile as soon as you stop looking. Earlier work on a related topic is "Statistically Rigorous Java Performance Evaluation", Georges, Buytaert, and Eeckhout [2007], which already discussed warmup-related problems but with an emphasis on the statistical analysis.
- The "SIGPLAN Empirical Evaluation Guidelines", Blackburn, Hauswirth, Berger, Hicks, and Krishnamurthi [2018]: This is recommended reading for SIGPLAN Artifact Evaluation Committee members, in particular the checklist which summariez a variety of good recommendations. The emphasis is on the things that you should *not* do when presenting benchmark results. My focus in the present document is to give recommendations on *how* you should proceed to produce benchmark results. The webpage also cites other interesting resources on benchmarking.

  Another resources with lists of mistakes to avoid include Gernot Heiser's list of System Benchmarking Crimes and Vitek and Kalibera [2012], which comes with interesting case studies and discussions of mistakes in published papers coming from incorrect methodologies.
- The course notes Moy [2021] provide an overview of many of these topics.

One hole in the literature in my opinion is an approachable introduction to benchmarking statistics. Fleming and Wallace [1986] provides a venerable introduction to this topic, but it only covers aspects which I find rather obvious: if you want to average ratios, use the geometric mean rather than the arithmetic means. Then some of the references include advanced discussions of p-values and student tests and ANOVA and what not – or just the advice to "ask a statistician" before putting ourselves in danger. I wonder, for example, if people can explain when it is better to use the "average" time rather than the "best" time to summarize benchmark results. The standard recommendation is to stick with "best" in general, and in my experience either of them lead to explainable results as long as the random noise is small enough.

## 3 The big picture: the Measure-Explain-Test-Improve loop

Let us assume that you want to compare a bunch of different implementations of a given software task (the *implementations*) on a bunch of different concrete instances (the *benchmarks*).

You have written your own scripts that, for each benchmark, run all the implementations enabled for this benchmark, collects the performance metrics you want to study (time, or memory, or the variation of time as a function of the input size, etc.), and presents the results in a readable, synthetic way (absolute or relative numbers, speedup ratios, graphs), etc.

The measure-explain-test loop then consists in repeatedly running the following steps:

**Measure** Run the benchmarks and produce results.

**Explain** Look at the results, and propose *explanations* for what you see. You want to explain the general, qualitative shape of the result (for example why implementation A does better than implementation B on a benchmark; or why C is so much worse), and in particular any specific thing that looks interesting or surprising (a bump in a curve, two results that are very close to each other while you would not have expected it, etc.).

Write down what you observe and your explanations. ("Implementation A is 1.5x-2.6x faster than B on inputs in range [a; b]. I think it comes from ...")

**Test** For each explanation that you proposed in the previous step, *test* it looking for concrete evidence for it.

For example, if you say "implementation A is faster than B because its variant of the function `foo` is faster", you can test this by using profiling tools on A and B to look at the respective time spent in `foo`, and checking that they indeed account for *most* of the observed difference. (If it accounts for *some* difference, but not most of it, then the explanation is wrong and needs to be revised into "A is faster than B because of `foo` and some other factor we don't understand yet".) If your explanation is that the compiler does a better job optimizing B than A, then do look at a dump of compiler output to check that the compiler indeed optimizes it better – and check that there are not other changes that could explain the difference. A good trick that you can use is to modify the implementation of A to have a *slower* `foo` function inspired by the one of B, and checking if the performance then become similar to B.

Write down what test you performed and what the results were.

**Improve** Perform any improvement ot the benchmarks or the implementations suggested by the previous steps.

- You will often find bugs in the benchmarks themselves, or at least ways to modify them to make their results less noisy, more reproductible, easier to explain.
- You may want to remove benchmarks that prove redundant (their explanations are too similar to another benchmark), or add new benchmarks suggested by your improved understanding.
- You may have found bugs or defects in the measured implementations themselves. Whenever you can you should try to fix them, or at least change your benchmarks to avoid unfair comparisons.

The iteration process stops when you feel confident that:

- you have tested explanations for all salient, qualitative aspects of the observed results,
- all benchmarks are interesting and useful, and
- you don't see important benchmarks to add to evaluate aspects that are within the scope of your research work.

You may also run out of time before you reach this fixpoint, with some parts of the results that you don't know how to explain, or for which you have plausible explanations that you were not able to test. These parts should be presented very clearly as unknown or conjectures in your presentation of the benchmark results.

*The wrong thing: half an iteration.* The common mistakes that benchmarking-beginners do is to perform only *half an iteration* of this METI loop: they write their benchmarks (testing them for the usual sort of correctness bugs of everyday programs: that they compute what you intend), run the measurements once (M), or maybe twice in a row to check that the results are stable, and then explain the results (E), and *that's it.* The explanations go in the paper. This is a terrible way to do benchmarking and the results are likely to contain *many* flaws. Some of those flaws may endanger the qualitative conclusions of the benchmark.

*Psychology of benchmarking.* One difficulty with generating good explanations is that we view the results with rose-tinted glasses, we already *know* that they are going to tell us that our favorite system is the fastest, and that it is the fastest specifically *because* of all our hard work. Testing our explanations rigorously prevents us from lying to ourselves. (But it does not protect against this bias completely, for example you can unvoluntarily bias the set of benchmarks, influenced by your knowledge of which aspect of the system you did optimize.)

## 4 Details

### 4.1 The parallel with testing

Software development often follows an Implement-Test-Fix (or Code-Test-Debug) loop, where you iterate testing your software until you believe that it correctly achieves its purpose (nowadays many agentic workflows are similar). It would be considered widely unreasonable to only do one-third of an iteration of the loop, implementing the code without testing it.

The purpose of a benchmark suite is not to compute a certain result correctly (which people typically test when they write benchmarks), but to give you a correct *understanding* of the performance of your program. The Measure-Explain-Test-Improve loop is a directly counterpart of the Implement-Test-Fix loop, where the things that are being tested is not the actual code of the benchmarks, but the performance analysis that you generate for their results.

This parallel between software testing and performance analysis testing goes a long way, and transfers to various finer-grained details of benchmarking.

### 4.2 Flavors of benchmarks

Some benchmarks are "microbenchmarks", they are simplistic/synthetic programs designed to evaluate a single aspect of your system, magnifying its performance impact. Some are "macrobenchmarks", they test a real-world workload, which typically exercize several different aspects of your system. Following the previous parallel, microbenchmarks correspond to unit tests, macrobenchmarks to integration tests.

I typically focus on microbenchmarks during the Measure-Explain-Test-Improve loop: by design they are easier to explain, test and improve. Then macrobenchmark act as sanity checks that our understanding of system performance applies not only to a few microbenchmarks curated to make our work look good, but also to realistic workloads.

In particular, looking at macrobenchmark results can reveal that your selection of microbenchmark is lacking, they do not suffice to explain the macro results. Trying to use your profiling/analysis expertise on the macrobenchmarks may suggest forgotten concerns and missing microbenchmarks.

### 4.3 Improving the implementations

The Improve step of the METI loop is described as improving the benchmarks. But often you will also think of ways to improve your program, not just the benchmarks: now that you are starting to have good benchmarks, you notice things to improve. Do not hesitate to do it! This is one of the very few benefits of doing your performance analysis very carefully.

Sometimes you will also notice ways to improve other implementations than your own – when trying to explain a performance difference, you will notice that it comes from a bug on their side that can be fixed. I also recommend doing this, it makes for a great story in the paper.

Finally, sometimes you will notice differences between implementations that make the benchmark results more complex than they should be; if you see a way to tweak the benchmark to control for these differences, I would recommend it: it simplifies the explanations and thus improves the benchmark.

### 4.4 Selecting implementations to compare against

Selecting implementations to compare against is a bit of an art. You don't need to compare the same implementations for all benchmarks – some choices only make sense for certain benchmarks. If that can inspire you, here are broad categories of implementations I have used in practice:

- Your competitors: the obvious idea is to compare against other software doing the same thing as you are doing, hoping to show that you improve over the state of the art (in some respects).
  (Note: if you expect to do better on some workloads and worse on others, do make sure to include benchmarks for both in your experiments. Including only the benchmarks were you expect to look good is dishonest, and it can hide bugs in your benchmark setup. For example maybe you are using incorrect build flags for your competitors making them artificially slower (forgetting to enable the "release profile" is a common one), so the comparison is broken. It is easy to miss those errors if you only compare on benchmarks where you already expect them to do worse.)
- A naive baseline: maybe there is a very simple way to do the thing you are doing, expected to be much slower. I recommend comparing it as well, you may be surprised by the results. (Or maybe your own code performs much worse than it should and you will detect it by comparing to this naive baseline.) For example, if you test parallel algorithms, the baseline would be a fast sequential implementation. If you test a sophisticated persistent data structure, the baseline would be a persistent map from the standard library.
- An *impossible* baseline: maybe there is a way for specific benchmarks to achieve the *perfect* implementation of a given system, to see how much faster it would be than whatever good work you are doing. For example, if you are trying to reduce FFI overhead, a cool trick is to write a benchmerk that can either be run through the FFI or with an equivalent-performance version not going through the FFI at all, so that you have the "zero FFI cost" baseline to compare against.
- The restricted competitor: sometimes your program solves a general problem, and some of your competitors only solve restricted versions of this problem. Consider writing benchmarks where the restricted version suffice, and compare your implementation against theirs – which can be assumed to be faster and simpler as it solves a simpelr problem.
- Unfair comparisons: maybe you are writing a program in a high-level programming language, and there exists a production implementation in carefully-optimized C. You know you are going to lose because of many different factors (many unrelated to the ideas you are developing in your work), but it is still good hygiene to include that unfair comparison in your benchmarks. (Be clear of course that the results are not directly comparable.)

## 5 After you are done: presenting the benchmarking results

Before presenting the results, you should write down:

- A short description of each benchmark, explaining which aspect of the system under test it is supposed to evaluate.
- A short description of each implementation.
  Provide URLs and mention which versions (and/or commit hashes) you are using. Right now it is obvious that the version of third-party library Foo that you are using is “the most recent ones”, but someone reading about your work in two years will have a hard time finding out which it is. (And no, people don’t want to download gigabytes of virtual machine images from Zenodo to find out.)

To present the results, you should carefully distinguish:

- *quantitative* results: “we observe speedup ranging from 1.6x to 2.7x”.
- *qualitative* results: “approach X is noticeably faster than Y, for example 2x faster in some cases”.

Quantitative results are precise, objective, and fragile to changes in benchmarking setups (different hardware, etc.). Qualitative results are less precise, and should be more robust.

(Notice that qualitative results can still contain ballpark numbers, “typically around 2x”. Producing some sort of ballpark estimate is important when you develop software that has users – they want to know if it will make a difference for them or not. I think that it is good hygiene in any case.)

A part of our job is to correctly translate the quantitative results we obtain into qualitative claims. You cannot expect someone that would reproduce the same experiment in a different environment to observe the same quantitative result (we could if we tested your work on all possible environements, but we do not have the money or time for this). But if they cannot reproduce your qualitative results, then your performance analysis was not good enough.

With this difference in mind, I try to do the following to present my benchmark results:

- In the introduction of the paper / when explaining the contributions, formulate a clear *qualitative* claim.
- In the benchmark-presentation section, repeat the qualitative claims.
- For each benchmark, present both quantitative results and a qualitative interpretation, and check that they agree with the global qualitative claims. (This part often ends up in an Appendix.) Give your explanations for the difference between implementations.

You don’t need to give *all the numbers* in the paper, typically just a few representative data points and/or well-chosen plots will do. (Even when detailed discussions are in Appendix, people like to see the plots in the paper if you have space.)

*Conjectures.* Sometimes a particular performance difference eludes you, you cannot find a good explanation. Or you have a plausible explanation, but were not able to validate it by testing. This is life, and you should be very clear about it in your presentation of the results. If you don’t know why a particular interesting phenomenon happens, just say so. If you were not able to test your hypothesis (some are very hard to test, for example locality effects), say it clearly.

We academics have very clear language we can use when we believe something, but have not validated it experimentally: “we conjecture that…”, or “our hypothesis is that …”. Use it.

## 6 Tips and tricks

(1) Each variant you are comparing should run several times, and you should observe the noise/variation between runs. If it is too high, there is probably a source of environmental noise that you should find out and eliminate.

 `hyperfine` is a *very* convenient tool to do this for you, but it can only measure runtime. I recommend checking it out.

(2) Every benchmark should come with a runtime parameter that varies the running time, that can be adjusted when running benchmarks on a slower or faster machine, or reduced to get faster feedback when running quick experiments to understand something.

 (I typically write my benchmarks as a single program that take command-line environments or environment variables to select the desired implementation, difficulty levels or other inputs, and an iteration count.)

(3) Make it easy to selectively enable only certain benchmarks, or tweak running times, this is also very useful to iterate faster on certain benchmarks and implementations.

 (Typically I do this by editing the scripts that run the benchmark. This is a bit dirty but trying to plan for variability with a complex set of command-line options is more work and less flexible. Do keep your benchmark scripts under version control so that you can see when you modified things. I like to also version the benchmark results data, so that anyone can clone the repository and analyze/plot things without having to rerun the benchmarks first.)

(4) Avoid comparing measurements from different runs of the benchmark, the risks of bias due to environmental noise are much larger that way.

 (For example: you recorded the results of for implementation A yesterday, and now you run implementation B and compare the results. But in fact you forgot to plug your laptop yesterday so the performance was worse and the results are incomparable, or you forgot that you modified the benchmark in the meantime, etc.)

(5) If you are using a laptop, do plug it to a power source and disable frequency-scaling. I use the Linux tool `cpupower frequency-info` to set my CPUs to a fixed, medium-low frequency (typically 2GHz) to avoid artifical slowdowns from heat management. You may want to look at whether your hardware has a "turbo" or "boost" mode and disable that.

 The good news is that if you forget to do this, it will typically *show* in the benchmark results as differences that you don't know to explain. For example a good sign of environmental noise is when you observe a clear performance trend over several iterations of the same test, for example they progressively get faster, or progressively get slower.

 You don't need to be paranoid about sources of environmental noise you don't know about, as long as you are rigorous in finding all "hmm, that's odd..." aspects of your benchmark results, and then in testing your explanations; following the process will find the noise sources that matter in your case.

(6) Look at the "extra time" spent by the benchmarked system outside the part that you are interested in. (For example: the "system" time of the process if you are benchmarking computations rather than OS interaction; or the time initializing the program, etc.) You need to tune the duration of your benchmark run so that this "extra time" is only a small fraction of the runtime, ideally close to the noise level you are willing to tolerate.

(7) I recommend recording your pre-benchmark expectations.

 Each careful benchmarking work I have participated in has generated surprises. When something is surprising at first, we work to understand it better, and after we do it is easy to minimize the surprise in retrospect and believe that we expected it all along. ("Gasp, third-party implementation X does better than us on benchmark C!" ⇒ "*Of course* our algorithm is not optimized for workloads such as C"). I like to write down in advance the qualitative results that I expect before running them; this forces me to be honest about the gap between my initial expectations and the actual results. I believe that it forces me to study things more carefully, and limits the influence of my natural bias towards my own work.

## 7 `hyperfine` demo

As mentioned earlier, `hyperfine` is a very convenient tool to run benchmarks. It works at the level of command-line invocations of processes: you provide commands and it measures the time it takes to run them, running it several times and warning about outliers. It also makes it very easy to compare the time of several different commands, with nice command-line output and markdown or JSON exports.

Let us compare two sorting functions, a quicksort and a mergesort on lists, implemented in OCaml. For this we write a program with the two implementations we want to compare, and use one or the other depending on an environment variable.

```
(* the implementations to compare *)
let rec quicksort = function
| [] -> []
| [x] -> [x]
| [x; y] as li -> if x < y then li else [y; x]
| x :: xs ->
  (* non-randomized pivot choices, apologies... *)
  let left, right = List.partition (fun y -> y <= x) xs in
  quicksort left @ [x] @ quicksort right

let rec mergesort = function
| [] -> []
| [x] -> [x]
| [x; y] as li -> if x < y then li else [y; x]
| li ->
  let rec split xs ys = function
  | z :: zs -> split (z :: ys) xs zs
  | [] -> (xs, ys)
  in
  let rec merge xs ys =
    match xs, ys with
    | zs, [] | [], zs -> zs
    | x::xs', y::ys' ->
      if x <= y
      then x :: merge xs' ys
      else y :: merge xs ys'
  in
  let (xs, ys) = split [] [] li in
  merge (mergesort xs) (mergesort ys)

(* command-line user interface;
   I never remember which arguments a given script expects,
   so I write the code to fail helpfully
    when they are missing or invalid. *)
let get_env var ~descr getter =
  let fail s = prerr_endline s; exit 2 in
  match Sys.getenv_opt var with
```

```
    | None -> Printf.ksprintf fail "environment variable %S is missing" var
    | Some str ->
      match getter str with
      | None ->
        Printf.ksprintf fail
          "environment variable %S has incorrect value %S; expected %s"
          var str descr
      | Some v -> v

let implementations = [("quicksort", quicksort); ("mergesort", mergesort)]

let impl_sort =
  get_env "IMPL" ~descr:"[quicksort | mergesort]" (fun s ->
    List.assoc_opt s implementations
  )

let size = get_env "SIZE" ~descr:"a number" int_of_string_opt
let n_iters = get_env "NITERS" ~descr:"a number" int_of_string_opt

(* running the benchmark *)
let input = List.init size (fun _ -> Random.int size)
let expected_output = List.sort Stdlib.compare input

let () =
  let output = impl_sort input in
  assert (output = expected_output);
  for _ = 1 to n_iters - 1 do
    ignore (impl_sort input)
  done
```

I compile this program (OCaml version 5.4, non-flambda native compiler) and look for parameter values that took between 200ms and 1s to run.

```
$ ocamlopt bench.ml -o /tmp/bench.opt
$ /tmp/bench.opt
environment variable "IMPL" is missing
$ time IMPL=quicksort NITERS=100 SIZE=10_000 /tmp/bench.opt

real    0m0.570s
user    0m0.553s
sys     0m0.012s
```

Just running `hyperfine` instead of `time` will run the command 10 times (by default) with nice progress bars and a summary of the result.

```
$ hyperfine "IMPL=quicksort NITERS=100 SIZE=10_000 /tmp/bench.opt"
Benchmark 1: IMPL=quicksort NITERS=100 SIZE=10_000 /tmp/bench.opt
  Time (mean ± σ):     641.2 ms ±  29.6 ms    [User: 622.2 ms, System: 9.8 ms]
  Range (min … max):   610.6 ms … 688.0 ms    10 runs
```

(Notice how the "System" time reported is small relative to the total runtime, well within the reported time interval. A "± 29.6 ms" variation is somewhat large, which may come from me running the benchmark on a train, with my computer unplugged.)

You can pass two different commands, and `hyperfine` will compare their running time.

```
$ hyperfine \
  "IMPL=quicksort NITERS=100 SIZE=10_000 /tmp/bench.opt" \
  "IMPL=mergesort NITERS=100 SIZE=10_000 /tmp/bench.opt"
Benchmark 1: IMPL=quicksort NITERS=100 SIZE=10_000 /tmp/bench.opt
  Time (mean ± σ):     608.6 ms ±  17.0 ms    [User: 592.7 ms, System: 9.0 ms]
  Range (min … max):   576.5 ms … 636.7 ms    10 runs

Benchmark 2: IMPL=mergesort NITERS=100 SIZE=10_000 /tmp/bench.opt
  Time (mean ± σ):     466.9 ms ±   7.6 ms    [User: 454.6 ms, System: 7.0 ms]
  Range (min … max):   454.1 ms … 476.1 ms    10 runs

Summary
  IMPL=mergesort NITERS=100 SIZE=10_000 /tmp/bench.opt ran
    1.30 ± 0.04 times faster than IMPL=quicksort NITERS=100 SIZE=10_000 /tmp/bench.opt
```

The option `-L <variable> <values>` will generate several commands by iterating on each value, it is very convenient to use for this sort of benchmarks where we control the command-line interface:

```
$ hyperfine -L impl quicksort,mergesort --command-name {impl} \
    "IMPL={impl} NITERS=100 SIZE=10_000 /tmp/bench.opt"
Benchmark 1: quicksort
  Time (mean ± σ):     611.3 ms ±  16.8 ms    [User: 593.1 ms, System: 10.6 ms]
  Range (min … max):   588.9 ms … 630.4 ms    10 runs

Benchmark 2: mergesort
  Time (mean ± σ):     467.2 ms ±   4.6 ms    [User: 453.4 ms, System: 8.5 ms]
  Range (min … max):   460.1 ms … 476.5 ms    10 runs

Summary
  mergesort ran
    1.31 ± 0.04 times faster than quicksort
```

I can also add `--export-markdown bench-data.md` to get a markdown-formatted array in addition. The results for yet another run:

```
| Command     |    Mean [ms] | Min [ms] | Max [ms] |    Relative |
|:------------|-------------:|---------:|---------:|------------:|
| `quicksort` | 622.6 ± 14.8 |    598.6 |    647.7 | 1.35 ± 0.04 |
| `mergesort` |  461.5 ± 5.6 |    453.9 |    469.2 |        1.00 |
```

Of course, this is only the "Measure" part of the METI iteration; now the real work would begin of trying to understand the observed differences, checking whether it holds for other input sizes, whether it comes from algorithmic differences or specificities of the inefficient list-based implementations, etc.

`hyperfine` is a very convenient tool that I would recommend including in your toolbox. Besides the aspect demonstrated above, it has the nice property of being portable across Linux, BSD and OSX, and even Windows. It has the following limitations (I have pondered writing my own command-line tool to lift them):

- `hyperfine` only measures time, not other metrics. I have occasionally wanted a tool to measure memory usage, for example.
- `hyperfine` only measures discrete data points, not the variation of times over a given input parameter. (If you try to use `-L` or other options to iterate on choices of input sizes, it will unhelpfully tell you that sort functions are much faster on smaller lists.)

I would like a similar tool that can measure input-interval-supported functions instead of single data points, and plot them conveniently. For the current time I use my own scripts to do this, running `hyperfine` to gather a stable mean for each data point, and then a bunch of ugly shell scripts to turn them into `gnuplot` inputs and render them.

## 8 Mistakes encoutered in the wild

Let us consider a few typical errors that can affect benchmark results, and how our methodology can help detect and fix them.

### 8.1 Two implementations are in fact the same

A surprisingly common error is to mess something up in benchmark scripts, in such a way that two implementations we are comparing are in fact the same. This can typically come from copy-paste errors at several possible places – the source code describing each implementation may be copied from one another and we forgot to make the necessary change, or the benchmark script may mistakenly call the same program with the same parameters twice, or the visualization script may ignore one dataset and plot from the same results twice.

When that happens the first reaction is to produce simple explanations for why the two implementations perform the same: "on this benchmark, the algorithmic differences do not matter in practice", etc. But if we are honest with ourselves (this is the key quality to produce good benchmarks), the results are *too* close to each other, it is surprising and this explanation must be tested carefully. For example: maybe there *is* an extreme input where you know that the implementations must behave differently, and you can tweak the benchmark to also measure it? Whatever change you make to either implementations to understand why they are performing so closely, you should notice that the results of *both* change when you modify one – and never change when you modify the other. You should catch the mistake before you manage to be fully convinced that the two implementations really perform exactly the same for this benchmark.

### 8.2 An implementation is wrongly fast, or wrongly slow

One of the implementations compared may be functionally incorrect in a way that makes it much faster than it should – it skips a part of the work. (One way to catch this earlier is to include reasonable assertions of functinoal correctness in the benchmarks directly. Be prepared to notice and report real bugs in some of the implementations you are comparing.) It can also be *wrongly slow*, for example if you use the wrong compilation options or more generally the wrong configuration. For example some languages have a "debug build" and a "relase build mode", with release-built programs several times faster; it is easy to jump to wrong conclusions if you test the competition in debug builds.

One way to detect these errors is to think, for each tested implementation, of a benchmark scenario where it should do very well, and test it. If it is wrongly slow, or if another implementation is wrongly fast, with a bit of luck you will get a counter-intuitive result that you will fail to explain.

Another way to detect this error is more direct but requires more domain expertise: you may notice that the speedup or slowdown faster is much higher than one would expect. Sometimes you have a good intuition for what sort of performance differences you should expect – 2x is good news, 10x is the sign of a bug. Sometimes you can make good estimates of the time that an implementatino should take: if it takes 20-50 cycles per operation, then it should take 1s-3s for the whole benchmark, so a result of 0.2s or 7s deserves careful testing.

A last way to detect compiler-related configuration errors is to *inspect the generated code* (in assembly language, or an intermediate representation of the compiler that is easier to understand yet should have had most optimizations applied). If the code is much worse than you would expect, there is a problem in the benchmarking setup.

### 8.3 An implementation is misused

A typical way to write comparative benchmarks is to define a common interface against which benchmarks are written, and write glue code to expose this interface from each implementation of interest. It is easy for mistakes to sneak in the glue code in a way that affects results and/or makes them more difficult to interpret.

For example, when comparing the performance of work-stealing schedulers, I was studying how performance varies with the number of worker threads. When compared the performance of three distinct schedulers, and realized that one had performance with $N$ workers that was comparable with the performance of the others at $N - 1$ workers – the graphs were effectively shifted sideways by 1, making it look slower. This came from a difference in interpretation of the worker-pool-creation parameters between libraries, which was not documented clearly.

At first we tried to explain the difference and take it into account in our comparison: "M looks slower, but its results at N=3 should be compared to the other at N=2, and then it is only 2% slower[...]". Eventually we changed the benchmark glue code to increment the worker-number parameter passed to that implementation, so that the results with $N$ workers would be comparable across implementations instead of requiring manual re-interpretation. This made our life (and our explanations) much easier and was obviously the right choice.

### 8.4 One benchmark has noisy result

"Noise" here refers to impact on benchmark results of factors that are unrelated to the evaluated implementations themselves, but other aspects of the computation environment.

Some sources of noise make results non-deterministic over several runs of the benchmark. If you follow the good practice of running your benchmarks several times and checking that the difference between runs remains small, you will detect it easily. If you don't, or if the benchmark has higher variations than expected, you are likely to detect it as you iterate on the benchmarks: you wrote down a description of the observed results on the previous iteration, and you now observe qualitatively different results for the same benchmark.

Other sources of noise are more dangerous because they come from aspects of your benchmark setup that do *not* typically change over several runs (for example your operating system). In one case we were measuring access times to a data structure (very short operations, typically of the order of a handful of reads from memory), and we could not explain why two implementations produced qualitatively different results, maybe a 10% performance gap, when they should have the same performance. After digging more we realized that this came from code-alignment issues, which we were able to test by generating 16 different copies of the benchmark binary with irrelevant constants added to change code alignment.

### 8.5 Irrelevant costs drown out the results.

It is good practice in your benchmarks to measure the time spent (or the other resources that you are using as benchmark metrics) in the core part of the implementation that you actually care about measuring, versus the time spent on other things: initial setup, disk access and other system calls, etc. You should tweak the benchmark code to ensure that most of the time is part in the "useful" part; this reduces environmental noise and makes benchmarks easier to interpret and explain.

For example the 'time' utility on Linux systems distinguishes "user" vs. "system" time, and you want the former to be noticeably larger than the latter. (Reduce the amount of IO operations if you can, or increase the difficulty parameter of the rest of the computation, etc.) Another source of "irrelevant cost" may come from the logic you implemented to run the same benchmark on several different implementations: the language abstractions involved can introduce indirections that drown out differences between implementations (when these differences are small, of the order of a handful of indirect function calls); it is good practice to check (via profiling and observation of the generated code) that the benchmark harness logic does not add noticeable costs.

## 9 Conclusion

A reasonable methodology to produce decent benchmarks is to run the following steps in a loop:

**Measure** Run the benchmarks and produce a good visualization of the results.

**Explain** Write explanations for every qualitative observation that can be made from the results.

**Test** Test each explanation experimentally to ensure that it actually explains most of the observed result.

**Improve** Improve everything that you have found to be wrong with the benchmarks, or distracts from what they are intended to measure.

At some point you will be satisfied that you have solid explanations for your benchmark results, and you will have fixed many errors or imperfections in the benchmarking setup. You can then select an interesting subset of quantitative results to provide to your readers, with their explanations. (If some part of the result remains unexplained or the explanation remains a conjecture, say so.) They should support a set of clearly-formulated qualitative claims.

## Bibliography

Philip J. Fleming and John J. Wallace. 1986. How not to lie with statistics: the correct way to summarize benchmark results. *Commun. ACM 29*, 3, 218–221. https://doi.org/10.1145/5666.5673.

Andy Georges, Dries Buytaert, and Lieven Eeckhout. 2007. Statistically rigorous java performance evaluation. *Proceedings of the 22nd Annual ACM SIGPLAN Conference on Object-Oriented Programming Systems, Languages and Applications*, Association for Computing Machinery, 57–76. https://doi.org/10.1145/1297027.1297033.

Todd Mytkowicz, Amer Diwan, Matthias Hauswirth, and Peter F. Sweeney. 2009. Producing wrong data without doing anything obviously wrong!. *Proceedings of the 14th International Conference on Architectural Support for Programming Languages and Operating Systems*, ACM, 265–276. https://doi.org/10.1145/1508244.1508275.

Jan Vitek and Tomas Kalibera. 2012. R3: repeatability, reproducibility and rigor. *SIGPLAN Not. 47*, 4a, 30–36. https://doi.org/10.1145/2442776.2442781.

Charlie Curtsinger and Emery D. Berger. 2013. STABILIZER: statistically sound performance evaluation. *Proceedings of the Eighteenth International Conference on Architectural Support for Programming Languages and Operating Systems*, Association for Computing Machinery, 219–228. https://doi.org/10.1145/2451116.2451141.

Edd Barrett, Carl Friedrich Bolz-Tereick, Rebecca Killick, Sarah Mount, and Laurence Tratt. 2017. Virtual machine warmup blows hot and cold. *Proc. ACM Program. Lang. 1*, OOPSLA. https://doi.org/10.1145/3133876.

Steve Blackburn, Matthias Hauswirth, Emery Berger, Michael Hicks, and Shriram Krishnamurthi. 2018. SIGPLAN Empirical Evaluation Guidelines. https://www.sigplan.org/Resources/EmpiricalEvaluation/.

Cameron Moy. 2021. How (Not) to Benchmark. https://felleisen.org/matthias/7480-s21/21.pdf.